\documentclass[aps,prl,preprint,groupedaddress,showpacs]{revtex4}

\usepackage[dvips]{graphicx}

\begin{document}

\title{Swapping and entangling qubits of single photons and atoms}

\author{T.~W.~Chen, C.~K.~Law and P.~T.~Leung}

\affiliation{Department of Physics, The Chinese University of Hong
Kong, Shatin, Hong Kong SAR, China}

\date{\today}

\begin{abstract}

A scheme is proposed here to achieve swapping and entangling of
photonic and atomic qubits with high fidelity. The mechanism is
based on the scattering of a single photon from a $\Lambda$-type
three-level atom. The evolution of the coupled system is analyzed
by projecting the quantum state onto a `bright' and a `dark'
state. Quantum interference of these two states, which is
determined by a frequency-dependent phase angle, can be exploited
to perform various two-qubit transformations. It is remarkable
that the probability of success of such transformations can
approach unity in the strong coupling cavity QED regime.

\end{abstract}


\maketitle

Quantum communication relies on the ability of transmitting and
retrieving quantum information at specified locations.
Particularly for a quantum network, communication between
spatially separated nodes requires inter-conversions between
flying qubits and stationary qubits with high fidelity. Therefore
it is important to investigate mechanisms that can accomplish the
task efficiently. Single photons and long-lived trapped atoms are
considered as fundamental hardware in distributed quantum
computing, with the former serving ideally as data bus, while the
latter playing the role of local quantum memory
\cite{Monroe-nature}. Indeed, various authors have proposed
protocols based on strong atom-field coupling in high-$Q$ cavities
\cite{Kimble-transfer,Haroche-memory}. These investigations
suggest that a complete transfer of a qubit from an atom to a
quantized electromagnetic field can in principle be feasible.
Recently, there are also studies addressing methods of quantum
communication using ensembles of atoms via their interactions with
light in free space
\cite{Fleischhauer,Communication-atoms-coherent-light}.

In addition to the task of quantum state transfer, it is often
desirable to {\em exchange} quantum information among carriers of
distinct nature \cite{Polzik-swapping}. A basic operation of this
kind is defined by the two-qubit transformation:
\begin{eqnarray}
&&  \left( {\alpha_1 \left| 1 \right\rangle _A + \alpha_0 \left| 0
\right\rangle _A } \right) \otimes \left( {c_1 \left| 1
\right\rangle _F + c_0 \left| 0 \right\rangle _F } \right)
\nonumber \\ && \to \left( {c_1 \left| 1 \right\rangle _A  + c_0
\left| 0 \right\rangle _A } \right) \otimes \left( {\alpha_1\left|
1 \right\rangle _F  + \alpha_0 \left| 0 \right\rangle _F }
\right), \label{swap-operation}
\end{eqnarray}
where $\alpha_1 \left| 1 \right\rangle _A + \alpha_0 \left| 0
\right\rangle _A$ represents an atomic qubit and  $c_1 \left| 1
\right\rangle _F  + c_0 \left| 0 \right\rangle _F$ denotes a
photonic qubit. Through the transformation~(\ref{swap-operation}),
the atomic qubit is deposited to the photon, and at the same time
the photonic qubit is mapped onto the atom. Such a swapping
process is more general than the typical one-way quantum state
transfer previously discussed in cavity QED systems, because the
latter corresponds to the special cases with either $c_1=0$ (atom
to photon) or $\alpha_1=0$ (photon to atom)
\cite{Kimble-transfer,Haroche-memory}. To our knowledge, physical
examples of swapping effects have only been investigated in atomic
ensembles with collective spin variables \cite{Polzik-swapping}
and NMR systems \cite{nmr}, a generic protocol of qubit swapping
among single photons and atoms in leaky optical cavities has not
yet been found.

In this Letter we present a mechanism to achieve qubit swapping
based on scattering of a single photon by an atom in a high-$Q$
cavity. The two ground states of a trapped $\Lambda$-type
three-level atom play the role of the two states of a stationary
qubit, and the two polarizations of a single photon constitute a
flying qubit. Quantum interference between the `bright' state and
the `dark' state of the $\Lambda$-system holds the key to
realization of qubit swapping. We derive explicitly an exact
analytic form of the scattering matrix, and determine a
frequency-dependent phase shift of the bright state. Such a phase
shift controls the quantum interference, and can be large in the
strong cavity coupling regime. Therefore the frequency degree of
freedom of photon would enable various kinds of two-qubit
transformations. In addition to the swapping
operation~(\ref{swap-operation}), we will also show how a
maximally entangled photon-atom pair can be generated at suitable
input photon frequencies.

\begin{figure}
\includegraphics[width=6.6cm]{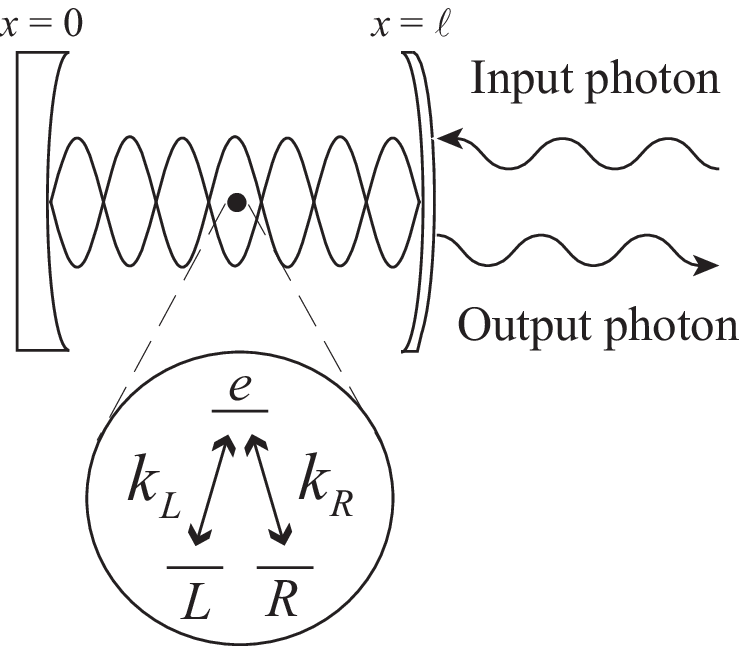}
\caption{A sketch of our cavity QED system.} \label{fig1}
\end{figure}

To begin with, we consider a one-dimensional cavity with length
$l$ and bounded by two mirrors. The left mirror at $x=0$ is
perfectly reflecting, while the other at $x=l$ is partially
transparent. The normal modes of the electromagnetic field are
characterized by a continuous wave number $k$. Specifically, the
spatial mode functions $u_k(x)$ are given by \cite{loudon}
\begin{equation}
\label{mode-function}
\label{fk} u_k(x)=\left\{\begin{array}{*{20}c} I(k)\sin{kx}\\
e^{-ikx}+R(k)e^{ikx} \end{array} \right.
\begin{array}{*{20}c}{}&{}&{}&{}&{}\\
{}&{}&{}&{}&{}
\end{array}
\begin{array}{*{20}c}
0&<&x&<&l \\ l&<&x&<&\infty
\end{array},
\end{equation}
where $I(k)=-2it/(1+re^{2ikl})$ and
$R(k)=(-r-t+re^{-2ikl})/(1+re^{2ikl})$, with $r$ and $t$ being the
reflection and transmission coefficients of the right mirror,
respectively \cite{mirror}. These continuous field modes provide a
basis for the field quantization. A $\Lambda$-type three-level
atom is located near the center of the cavity. The atom has two
degenerate ground states $|L\rangle$, $|R\rangle$ and an excited
state $|e\rangle$. The ground state $|L\rangle$ ($|R\rangle$) can
be excited to $|e\rangle$ by absorbing a $|k_{L}\rangle$
($|k_{R}\rangle$) mode photon as shown in Fig.~\ref{fig1}, where
the subscripts $L$, $R$ denote the polarization of the photon. In
our scheme, there is one and only one photon involved in the
scattering process, and no external classical pump fields are
required. To facilitate our discussion, we first neglect the
spontaneous emission from the atomic excited state into side modes
of the cavity. The loss due to spontaneous decay will be discussed
later in the paper.

The Hamiltonian of our model (in units of $\hbar=c=1$) is given by
\begin{eqnarray}
\label{H} \hat{H} &=& \omega _{e}|e\rangle \langle
e|+\int_0^\infty dk \, \sum_{\mu=L,R}k \hat a_{k\mu}^{\dag}\hat
a_{k\mu} \nonumber \\ && + \int_0^\infty dk \,
\sum_{\mu=L,R}g_{\mu}(k)\hat a_{k\mu}|e\rangle \langle \mu|+{\rm
h} .{\rm c}..
\end{eqnarray}
Here, $\hat a_{k\mu}$ and $\hat a_{k\mu}^{\dag}$ are the
annihilation and creation operators associated with the field mode
$u_k(x)$ with the polarization $\mu$ ($\mu = L, R$). The dipole
coupling strength $g_{\mu}(k)$ is given by
\begin{equation}
g_{\mu}(k)=\frac{\lambda_{\mu}\sqrt{\kappa/\pi}e^{i\theta_{\mu}}}{
k-k_c+i\kappa}, \label{gmuk}
\end{equation}
which is proportional to the mode strength at the location of the
atom, i.e., $u_k(x=l/2)$. The $\kappa = -\ln |r| /2l$ is the
leakage rate of the cavity,
$\lambda_{\mu}^2=\int_{-\infty}^{\infty}|g_{\mu}(k)|^2dk$, and
$\theta_{\mu}$ is a phase angle associated with dipole transition
matrix elements. In writing Eq.~(\ref{gmuk}), we have assumed that
only one of the cavity quasi-modes (with a resonance frequency
$k_c$ defined by the real part of the pole of $R(k)$) interacts
with the atom. Such an approximation is valid when $\omega_e
\approx k_c$, and all other quasi-modes are far off resonance.
This requires a small cavity with high finesse in typical cavity
QED experiments \cite{Kimble-logic-gate}. Note also that in the
optical regime, the approximation allows us to extend the lower
bound of the frequency of the photon from $0$ to $-\infty$.

Initially, the atom is prepared in a coherent superposition of the
two ground states and a single photon wave packet of arbitrary
spectrum and polarizations is injected into the cavity. Our task
is to determine the analytic solution of the final state of the
system. First, we notice that the scattering process is energy
conserving. Hence for a given $k$, the initial and final states
dwell in the same $k$-subspace spanned by the four basis vectors:
$|L; k_L\rangle$, $|L;k_R\rangle$, $|R;k_L\rangle$, and
$|R;k_R\rangle$. It is obvious that $|L;k_R\rangle$ and
$|R;k_L\rangle$ are eigenvectors of $\hat H$. In addition, the
coherent superposition ${g_R(k) |L; k_L\rangle - g_L(k) |R;
k_R\rangle}$ is a dark state well known in $\Lambda$-systems. The
excited state $|e \rangle$ couples only with the `bright' state
\begin{equation}
\label{bright} |\psi(k) \rangle \equiv \frac{1}{V(k)} \left[
{{g_L^*(k) |L; k_L\rangle + g_R^*(k) |R; k_R\rangle} }\right].
\end{equation}
where $V(k)=\sqrt{|g_L(k)|^2+|g_R(k)|^2}$. Therefore we only need
to solve the evolution of $|\psi(k) \rangle$, which is governed by
the Hamiltonian~(\ref{H}) in the corresponding subspace:
\begin{eqnarray}
\label{H'} \hat{H}'&=& \omega _{e}|e; \phi \rangle \langle e;
\phi|+\int_{-\infty}^\infty dk \, k |\psi(k)\rangle\langle\psi(k)|
\nonumber \\ && + \int_{-\infty}^\infty dk \, V(k)|e; \phi
\rangle\langle\psi(k)|+{\rm h} .{\rm c}.,
\end{eqnarray}
with $\phi$ being the vacuum of the field.

For the scattering process considered here, we are interested in
the transition matrix element $U_{kk^{\prime}} = {\langle\psi(k)|}
\hat U(t) |\psi(k^{\prime}) \rangle$ in the asymptotic long time
limit, where $\hat U(t)$ is the evolution operator. With the help
of the resolvent formalism \cite{Cohen}, we obtain the relation:
\begin{equation}
\label{S-matrix} U_{kk^{\prime}}=\delta(k-k^{\prime})\left[1-2\pi
i V(k)^2 \langle e;\phi|\hat{G}(k)|e; \phi \rangle\right]e^{-ikt},
\end{equation}
where $\hat{G}(k) \equiv (k-\hat{H}')^{-1}$ is the resolvent. The
element $\langle e; \phi |\hat{G}(k)|e; \phi \rangle$ can be
determined by the projection operator method, which gives
\begin{equation}
\label{Gee} \langle e; \phi |\hat{G}(k)|e; \phi \rangle
=\frac{\Delta k+i\kappa}{(\Delta k-\omega_+)(\Delta k-\omega_-)}.
\end{equation}
Here $\Delta k=k-k_c$, $\omega_{\pm}=(\delta_e-i\kappa)/2\pm
\sqrt{\left({\delta_e
+i\kappa}\right)^{2}/4+\lambda_L^2+\lambda_R^2}$ is related to
vacuum Rabi-splitting \cite{Kimble_splitting}, and
$\delta_e=\omega_e-k_c$. Together with Eq.~(\ref{S-matrix}), we
have $\hat U(t,0)\left| {\psi (k)} \right\rangle =e^{-ikt}
e^{i\delta _s (k)} \left| {\psi (k)} \right\rangle$. The exact
expression of the phase shift $\delta_s(k)$ is given by
\begin{equation}
\label{phase} e^{i\delta_s(k)}=\frac{(\Delta k-\delta_e)(\Delta
k^2+\kappa^2)-(\Delta k+i\kappa)(\lambda_L^2+\lambda_R^2)}{(\Delta
k-\delta_e)(\Delta k^2+\kappa^2)-(\Delta
k-i\kappa)(\lambda_L^2+\lambda_R^2)}.
\end{equation}

Let us now denote a general input state (at a given $k$) by $|{\rm
in};~k\rangle =
\alpha_1(k)|L;k_L\rangle+\alpha_2(k)|R;k_R\rangle+\alpha_3(k)
|L;k_R\rangle+\alpha_4(k)|R;k_L\rangle$, and the corresponding
output state by $|{\rm out};~k \rangle =
\beta_1(k)|L;k_L\rangle+\beta_2(k)|R;k_R\rangle+\beta_3(k)
|L;k_R\rangle+\beta_4(k)|R;k_L\rangle$. With the help of
Eq.~(\ref{phase}) and taking account of the (non-interacting) dark
state, the input-output relation can be conveniently expressed in
a matrix equation,
\begin{equation}
\label{Transform} \left(\begin{array}{c}  \beta_1\\ \beta_2
\\ \beta_3
\\ \beta_4
\end{array}\right)
=\left(\begin{array}{cccc} T_{LL}  & T_{LR} & 0 & 0 \\ T_{RL} &
T_{RR} & 0 & 0 \\ 0  & 0 & 1 & 0 \\ 0  & 0 & 0 & 1
\end{array}\right)
\left(\begin{array}{c} \alpha_1  \\ \alpha_2  \\ \alpha_3
\\ \alpha_4
\end{array}
\right).
\end{equation}
Here $T_{LR}(k)=T_{RL}(k)e^{2i(\theta_R-\theta_L)}
={g^\ast_L(k)g_R(k)}(e^{i\delta_s(k)}-1)/{V(k)^2}$,
$T_{LL}(k)={e^{i\delta_s(k)}(|g_L(k)|^2+|g_R(k)|^2)}/{V(k)^2}$,
$T_{RR}(k)={e^{i\delta_s(k)}(|g_R(k)|^2+|g_L(k)|^2)}/{V(k)^2}$,
and we have omitted the trivial free evolution phase factor
$e^{-ikt}$.

Equation~(\ref{Transform}) describes a general transformation of
photon-atom states before and after the scattering process. The
matrix elements $T_{ij}(k)$ $(i,j = L,R)$ are the consequences of
interference between amplitudes associated with the dark state and
the bright state $|\psi (k) \rangle$ that acquires a phase shift
$\delta_s(k)$. Notice that the elements $T_{ij}(k)$ form a
sub-matrix that is unitary. By tuning the photon frequency and
cavity parameters, we are able to perform qubit transformations
between the atom and the photon. In the following, we consider
swapping and entangling of qubits.

\vspace{5mm} \noindent {\em Swapping qubits}--- If we treat the
labels $L(k_L)$ and $R(k_R)$ as logical $1(0)$ and $0(1)$ for the
atom(photon), respectively, the qubit swapping
operation~(\ref{swap-operation}) corresponds to the conditions:
$T_{LL}=T_{RR}=0$ and $T_{LR}=T_{RL}=1$. Such conditions are
satisfied if $g_L(k)=-g_R(k)$ and
\begin{equation}
\delta_{s}(k)= \pm \pi.
\end{equation}
The condition $g_L(k)=-g_R(k)$ can be satisfied by choosing
suitable atomic transition schemes. For example, the D1 line of
sodium with hyperfine ground states $|L \rangle = |F=1, m_F=-1
\rangle$ and $|R \rangle = |F=1, m_F=1 \rangle$, and the excited
state $|e \rangle = |F=1, m_F=0 \rangle$ is a candidate of
$\Lambda$-systems with equal but opposite dipole matrix elements.

If the cavity is tuned at the resonance $k_c=\omega_e$, the
requirement of a $\pi$ phase shift $\delta_s(k)$ is satisfied for
the photon frequencies:
\begin{equation}
k=k_c, k_c\pm\sqrt{2\lambda^2-\kappa^2}, \label{3-roots}
\end{equation}
where $\lambda\equiv \lambda_L = \lambda_R$. Hence, if the
frequency of the incident photon is given by Eq.~(\ref{3-roots}),
perfect qubit swapping can be achieved as long as the loss due to
spontaneous decay is ignorable.

Furthermore, we can extend our analysis to incorporate a non-zero
spontaneous decay rate $\gamma$ of the excited state. This can be
done by adding a negative imaginary part $-i\gamma$ to the atomic
frequency $\omega_e$. Accordingly, Eq.~(\ref{phase}) is modified
but the general form of input-output transformation given by
Eqs.~(\ref{Transform}) remains unchanged. We find that among the
three characteristic frequencies given in Eq.~(\ref{3-roots}),
$k=k_c$ provides a robust performance against spontaneous emission
loss in the strong coupling regime, where $\lambda^2 \gg \kappa
\gamma$ and consequently $e^{i\delta_s(k_c)} \approx -1 + \kappa
\gamma / \lambda^2$. Therefore the loss due to spontaneous decay
is of order $\kappa \gamma / \lambda^2 $, and becomes
insignificant in this regime.

To complete our analysis we need to consider realistic photons in
forms of wavepackets, instead of being purely monochromatic light.
We consider an initial state $|\Phi_{\rm
in}\rangle=\left(A_L|L\rangle+A_R|R\rangle\right)
\otimes\int_{-\infty}^{\infty}dk \, f_S(k)\left[ { C_L \left| k_L
\right\rangle  + C_R \left| k_R \right\rangle } \right]$. To apply
our solution (7)-(10), the spectral function $f_S(k)$ of the
photon packet should be taken such that the atom does not
experience the field of the injected photon for $t \le 0$. For
concreteness, we consider a Gaussian photon packet which is
traveling towards the cavity from a far distance $x_0$ at $t=0$
with a peak frequency $k_c$ and spectral width $\kappa_{\rm in}$
\cite{ourpra}:
\begin{equation}
\label{input-gaussian} f_S(k^{\prime
})=\frac{1}{\pi^{1/4}\sqrt{\kappa_{\rm in}}}\,
\exp\left[-\frac{(k^{\prime}-k_c)^2}{2\kappa_{\rm
in}^2}+ik^{\prime}x_0\right].
\end{equation}
In the asymptotic long time limit, the output state $|\Phi_{\rm
out}\rangle$ is determined by the transformation~(\ref{Transform})
for each $k$. The swapping fidelity is defined by the overlap
$F=|\left\langle\Phi_{\rm swap}|\Phi_{\rm out}\right\rangle |^2$,
where $|\Phi_{\rm
swap}\rangle=\left(C_R|L\rangle+C_L|R\rangle\right)
\otimes\int_{-\infty}^{\infty}dk \, f_S(k)\left[ { A_R \left| k_L
\right\rangle  + A_L \left| k_R \right\rangle } \right]$ is the
ideal swapping of the input state.

After some straightforward calculations, we have $F=1-2{\rm
Re}(\xi)\eta+|\xi|^2\eta^2$, where $\eta=|A_LC_L-A_RC_R|^2$ and
$\xi=\int_{-\infty}^{\infty}T_{LL}(k)|f_S(k)|^2dk$. Therefore the
fidelity depends on two parameters $\eta$ and $\xi$, which in turn
depend on the qubit states and the photon spectrum. In essence
$\eta$ measures the overlap of the initial state with the
interacting state $|\psi(k) \rangle$ defined in
Eq.~(\ref{bright}), and $\xi$ measures the deviations of $T_{ij}$
from the ideal swapping matrix elements $T_{LL}=0$ and $T_{LR}=1$
averaged over the incident photon spectrum.

\begin{figure}
\includegraphics[width=6.6cm]{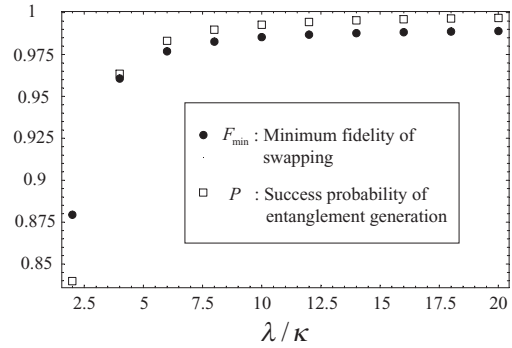}
\caption{Dependence of minimum fidelity of swapping (filled
circles) and success probability of Bell's state generation
(squares) on the normalized coupling strength $\lambda/\kappa$ for
$\gamma=0.5\kappa$ and $\kappa_{\rm in}=0.1\kappa$.} \label{fig2}
\end{figure}

Further calculations show that $F$ attains minimum when $\eta=1$
for any $\xi$, therefore $F$ obeys the inequality,
\begin{equation}
F\ge1-2{\rm Re}(\xi)+|\xi|^2=F_{\rm min}.
\end{equation}
To provide numerical examples, the circles in Fig.~\ref{fig2} show
the dependence of $F_{\rm min}$ on the normalized coupling
strength $\lambda/\kappa$. In the case $\lambda/\kappa=10$,
$\gamma/\kappa=0.5$, we have $F_{\rm min} \approx 97\%$ for a
Gaussian packet $\kappa_{in}=0.1 \kappa$. Higher $F_{\rm min}$ can
be achieved by decreasing the spectral width of the input photon.

\vspace{5mm} \noindent {\em Entangling qubits} --- If
$T_{RL}(k)=T_{LL}(k)e^{i\theta}$ (where $\theta$ is a real phase
angle) is satisfied at certain frequencies, then an initial
product state $|L,k_L \rangle$ at those frequencies will evolve
into a Bell's state: $|\Phi_E \rangle = (|L,k_L \rangle + e^{i
\theta} |R,k_R \rangle ) / \sqrt{2}$. In other words, the
scattered photon and the final state of the atom can become
maximally entangled.

We find that solutions of $k$ satisfying $T_{RL}(k)=T_{LL}(k) e^{i
\theta}$ do exist. Under the condition $\delta_e =0$,  we have
$\Delta k = \pm \kappa$ (or $k=k_c \pm \kappa$) as approximate
solutions in the strong coupling regime $\lambda \gg \gamma,\
\kappa$, and the corresponding $\theta$ are $\mp \pi/2$. Such
photon frequencies do not depend on the coupling strengths, hence
the entangling operation is not sensitive to the position of the
atom as long as the strong coupling requirement is satisfied.

To discuss the success probability, let us consider an initial
state $|\Phi_{\rm
in}\rangle=|L\rangle\otimes\int_{-\infty}^{\infty}dk\,f_E(k)
|k_L\rangle$, where $f_E(k)$ peaks at $k=k_c+\kappa$ with a
spectral width $\kappa_{\rm in}$. The success probability is
defined by $P=\left|\left\langle\Phi_E|\Phi_{\rm
out}\right\rangle\right|^2$, where $|\Phi_E \rangle =
\int_{-\infty}^{\infty}dk\,f_E(k)(|L,k_L \rangle -i |R,k_R \rangle
) / \sqrt{2}$ is the target Bell's state. It can be shown that:
$P=\frac{1}{2}+|\xi|^2- {\rm Re}(\xi)+{\rm Im}(\xi)$, where $\xi$
is similarly defined by the previous expression, with $f_S(k)$
replaced by $f_E(k)$. It is useful to remark that $P$ depends on
the photon spectrum $|f_E(k)|^2$. The phase of $f_E(k)$, which
defines the spatial details of the photon wavefunction, does not
affect the success probability. The same is also true for the
swapping process above. We find that the probability of generating
the Bell's state is quite high in the strong coupling regime as
long as the spectral width of the incident photon is sufficiently
narrow. Figure~\ref{fig2} illustrates how $P$ depends on the
normalized coupling strength $\lambda/\kappa$. For example at
$\lambda / \kappa =10$, $P\approx 99.2\%$ can be achieved by using
$\gamma=0.5\kappa$ and a Gaussian amplitudes $f_E(k)$ with
$\kappa_{\rm in}=0.1\kappa$.

To conclude, we have demonstrated how a single photon and an atom
can effectively `communicate' with each other through scattering
in a strong coupling cavity QED environment. An interesting aspect
of our approach is that the frequency of the photon serves as a
control parameter. Various qubit transformations can be achieved
in a tunable way according to the scattering
matrix~(\ref{Transform}). The continuous dependence of frequencies
in such matrix not only allows swapping and entangling in the same
setup, but also enables the preparation of a wide range of
photon-atom states when combining local transformations of the
atom and the photon individually. Although we have used a
one-dimensional cavity to illustrate the process, the
Hamiltonian~(\ref{H}) itself is quite general. Our method is
equally applicable to systems in higher dimensions as long as a
single quasi-mode and a dominant input-output channel is involved
in the process. In addition, since our protocol concerns only
asymptotic states,  no precise timing of interaction period is
required as compared with typical state control strategies
\cite{Swap_remark}. Finally, we note that the realization of our
scheme relies on deterministic single photon sources
\cite{Law-Kimble,Rempe-photon_source} and trapping of single atoms
inside optical cavities \cite{Mundt,Mckeever}. Our work in fact
addresses a novel application of single photons scattering once
these technologies become available.

\vspace{5mm}

\begin{acknowledgments}
We thank H.T. Fung for useful discussions. Our work is supported
in part by the Hong Kong Research Grants Council (grant No.
CUHK4282/00P and CUHK4016/03P).
\end{acknowledgments}


\begin{references}
\bibitem{Swap_remark} According to our time-evolution calculations,
a time with several $\kappa_{in}^{-1}$ is sufficient for the
system to reach asymptotic states, when $\kappa_{in}$ is the
smallest parameter among $\kappa$, $\gamma$ and $\lambda$.
\end{references}
\end{document}